\documentstyle[pra,aps,floats,psfig]{revtex}

\draft

\begin{document}

\title{Quantum and Semiclassical Calculations of Cold Atom Collisions in
Light Fields}

\author{K.-A. Suominen}
\address{Helsinki Institute of Physics, PL 9, FIN-00014 Helsingin yliopisto, 
Finland}

\author{Y. B. Band and I. Tuvi}
\address{Departments of Chemistry and Physics, Ben-Gurion University,  Beer
Sheva 84105, Israel}

\author{K. Burnett}
\address{Clarendon Laboratory, Department of Physics, University of Oxford,
Parks Road, Oxford OX1 3PU, United Kingdom}

\author{P. S. Julienne}
\address{Atomic Physics Division, National Institute of Standards and
Technology, Gaithersburg, MD 20899}

\maketitle

\begin{abstract}
We derive and apply an optical Bloch equation (OBE) model for describing
collisions of ground and excited laser cooled alkali atoms in the presence
of near-resonant light. Typically these collisions lead to loss of atoms
from traps. We compare the results obtained with a quantum mechanical complex 
potential treatment, semiclassical Landau-Zener models with decay, and a 
quantum time-dependent Monte-Carlo wave packet (MCWP) calculation. 
We formulate the
OBE method in both adiabatic and diabatic representations. We calculate the
laser intensity dependence of collision probabilities and find that the
adiabatic OBE results agree quantitatively with those of the MCWP calculation,
and qualitatively with the semiclassical Landau-Zener model with delayed
decay, but that the complex potential method or the traditional Landau-Zener
model fail in the saturation limit. 
\end{abstract}


\section{Introduction}\label{intro}

Collision dynamics of cold atoms in laser traps have been extensively
investigated over the past few years. When the red detuning $\Delta$ from
the atomic resonance frequency is large compared to the natural decay rate
$\gamma$, we find a photoassociation spectrum of isolated bound 
vibrational-rotational levels in the attractive excited state potentials.
This is now a highly developed subject and is fairly well understood, see 
the review~\cite{PAreview}. In contrast, when
$\Delta$ is small, on the order of $\gamma$, the mechanisms and rate 
coefficients of trap loss processes which result from photoexcitation of the
diatomic quasimolecule at long range are still rather poorly understood
theoretically~\cite{Smith,Suominen}, in spite of the numerous experimental
studies of this subject~\cite{experiments}.

The reason for this is twofold: real hyperfine structure introduces much
complexity into the collision dynamics, and the prominent role of  excited
state spontaneous decay during the very long time scale of the collision is
difficult to calculate quantum mechanically.  The number of degrees of
freedom associated with the spontaneous emission is, of course, infinite.  
Adiabatically
eliminating these degrees of freedom leads to a mixed state representation
that can not be described in terms of wavefunctions but requires solving the
Liouville-von Neumann equation for the quantum mechanical density matrix
$\rho(R,R';t)$~\cite{densmat}:
\begin{equation}
   \frac{\partial}{\partial t} \rho(R,R';t) = -\frac{i}{\hbar} \left[
   H(R)\rho(R,R';t)- \rho(R,R';t) H(R')\right]  + \Gamma \rho(R,R';t) ,
   \label{liouville1}
\end{equation}
where $H(R) = T(R) + V(R)$ is the system Hamiltonian for kinetic energy $T(R)$ 
and interaction potential $V(R)$, and $\Gamma$ is the decay tensor.  Thus, the
theoretical treatment of cold atom collisions  serves as both prototype and
paradigm for new constructs to treat non-equilibrium open systems  coupled to
reservoirs.  Since the direct solution of Eq.~(\ref{liouville1}) for cold
collision situations is beyond currently available computational
resources~\cite{Lai93,SG93}, approximate methods for treating the collision
dynamics in light fields must be developed.

The methods currently available are the semiclassical local equilibrium model
of Gallagher and Pritchard~\cite{GP} or Julienne and Vigu\'e~\cite{JV},
the semiclassical dynamical Landau-Zener
models~\cite{JSB,HSB94,Suominen94,Suominen95}, the semiclassical optical Bloch
equation (OBE) method~\cite{BJ}, the quantum complex potential
method~\cite{JSB,TB,BVT}, and the Monte Carlo wave packet method of simulating
the full quantum density matrix~\cite{HSB94,Suominen94,Suominen95}.  Although
the latter is capable in principle of treating the full quantum dissipative
dynamics for an arbitrary number of coupled states in arbitrarily strong
laser fields, the method is extremely computer intensive, and therefore slow 
and impractical. The complex potential method can treat fast many
coupled channels quantum mechanically, including bound state resonances, but
only in the limit of very weak laser fields where no more than one excitation
and decay event per collision occurs.  

The semiclassical methods are very appealing because of their computational 
tractability, simple interpretation and physical picture of the collision. 
However, several fully quantum calculations~\cite{HSB94,Suominen94,Suominen95}
have shown that both the local equilibrium and semiclassical OBE methods (in 
the formulation given in Ref.~\cite{BJ}) give incorrect results by an order 
of magnitude or more for detunings of a few $\gamma$ or less for temperature 
$T < 1$  mK, depending on species.  So
far no practical theory exists for ultracold collisions for realistic atoms in
a light field with $\Delta\simeq\gamma$ which is fully quantum mechanical and
also capable of treating dissipation and decay.  Therefore, there is not yet
any satisfactory description of trap loss rates in the small detuning limit.
For large detunings, $\Delta\gg\gamma$, collision in a light field goes to
photoassociation spectroscopy, in which isolated molecular bound vibrational 
levels are excited. Resonant scattering theory does then an excellent job 
of explaining the excitation rate~\cite{PAreview,Reginaldo}.

Quantum calculations have shown that semiclassical methods may
still be useful in characterizing cold collisions in a light field with
$\Delta\simeq\gamma$~\cite{HSB94,Suominen94,Suominen95}. The local equilibrium 
model for cold collisions place a prominent emphasis on off-resonant 
quasimolecular excitation outside the region around the Condon point 
$R_C$. In contrast to them, a semiclassical picture based on localized 
Landau-Zener excitation near $R_C$ with subsequent semiclassical evolution 
with decay inside $R_C$ gives an excellent representation of the quantum 
dynamics for $T$ near 1 mK, when compared with results obtained from 
quantum mechanical calculations. The Landau-Zener model only begins to
fail near $T = 1\ \mu$K and at large intensities~\cite{Suominen94}. Therefore,
since there still seems to be good opportunities for semiclassical models, we 
have revisited the OBE method, and provided a rigorous derivation of the 
velocity-corrected semiclassical OBE equations starting with the appropriate
quantum mechanical equations (replacing the unsatisfactory approach used in
Ref.~\cite{BJ}).  We show that an {\em adiabatic} rather than a {\em diabatic}
formulation of the semiclassical OBE equations gives quite good agreement
with the quantum methods, even at saturation, in contrast to the poor
agreement provided by the diabatic treatment.  

We find that the adiabatic OBE calculations are in good agreement with the
MCWP results and time-independent complex potential results  (used for weak
laser fields where this method is appropriate), down to low collision
temperatures. Only upon increasing the laser detuning above the onset of
resonances due to bound state structure does the adiabatic OBE method
fail~\cite{JSB}.  We also use an improved Landau-Zener model with
dissipation~\cite{Suominen95} which works even in the presence of strong
saturation, where the complex potential method  fails. This model offers a
qualitative understanding of the strong field processes. The numerical
comparisons are for the standard two-state model on which the quantum and
other semiclassical models have heretofore been tested.  Although these test
calculations ignore the complex multistate structure introduced by molecular
hyperfine structure, the hope is that semiclassical methods can yet be
developed that are capable of treating the complexity of multistate collision
dynamics in the presence of decay. 

This paper is constructed as follows. Section~\ref{model} presents the model of
trap loss processes we shall  use in order to test the methods developed and
employed.  Section~\ref{obe} contains the derivation of the OBE method using
the two different bases.  Section~\ref{MC} describes the Monte-Carlo method
which serves as the standard against which the approximate methods we use are
compared.  Section~\ref{complex} describes the complex potential method. 
Section~\ref{LZ} develops the generalized Landau-Zener approach to strong
laser field cold atom collisions.  Section~\ref{results} presents the
comparison of the numerical results from the various methods, and
Sec.~\ref{final} contains a summary and conclusion.

\section{The quasimolecule model for cold collisions}\label{model}

\begin{figure}
\centerline{\psfig{width=120mm,file=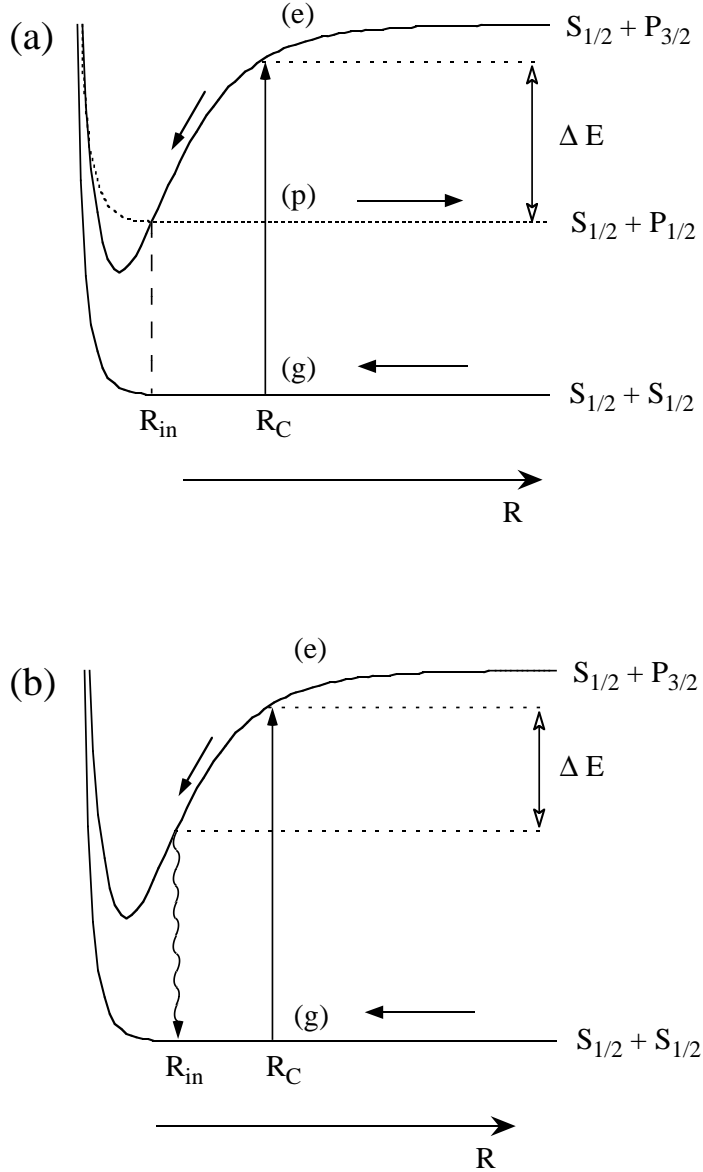}}
\caption[f1]{The basic trap loss mechanisms. The figures show the
quasimolecule potentials as functions of the internuclear separation $R$. The
corresponding asymptotic combinations of the atomic states are also given. In 
(a) we demonstrate the fine-structure change (FS) loss mechanism. The 
quasimolecule is excited from the ground state $g$ to the state $e$, then it 
moves towards small $R$, where it is transferred to the $p$ state at $R=
R_{\rm in}$. Finally the atoms exit the collision on this state, sharing a 
kinetic energy increase equal to $\Delta E$. In (b) the excited state $e$ 
decays back to the ground state (g) after the atoms have gained enough kinetic 
energy ($\Delta E$) to escape from the trap. This is the radiative escape 
mechanism (RE). If decay takes place too soon, i.e., at $R>R_{\rm in}$, then 
the escape turns into heating due to insufficient increase in the kinetic 
energy (the trap depth is not exceeded).
\label{lossmech}}
\end{figure}

The basic loss processes for atom traps due to cold collisions are
fine-structure change (FS) and radiative escape (RE)~\cite{GP,JV}; these
are demonstrated in Fig.~\ref{lossmech}. Due to the low temperatures we can
consider the collision of two atoms as internal dynamics of a diatomic
quasimolecule. The simplest prototype model for the FS mechanism consists of
three collision channels, i.e., quasimolecule states. The model described here
is identical to that used in Ref.~\cite{Rapid}.  We ignore any rotational
structure, i.e., only the s-wave is considered, but the model can be extended 
to higher partial waves. Here the three
channels are: the ground $^2S_{1/2}$ + $^2S_{1/2}$ state channel labelled $g$,
the excited $^2S_{1/2}$ +  $^2P_{3/2}$ state channel labelled $e$ and a probe
channel (correlating asymptotically with $^2S_{1/2}$ + $^2P_{1/2}$ state)
labelled $p$. 

In the FS mechanism the system starts on channel $g$, and is later excited at
the Condon point $R_C$ to the channel $e$, which has an attractive potential.
When the atom reaches the crossing between the potentials for the $e$ and $p$
channels, it may enter the $p$ channel and eventually come out of the
collision having gained as kinetic energy the energy difference between the
$^2S_{1/2}$ + $^2P_{3/2}$ and $^2S_{1/2}$ + $^2P_{1/2}$ states. This gain is
large enough to propel atoms from the shallow trap. 

In the RE mechanism the quasimolecule decays back from the $e$ channel to the
$g$ channel via spontaneous emission. If this decay does not take place too
early, the acceleration on channel $e$ will give the atoms enough kinetic
energy to escape from the trap. 

For both mechanisms we need to find out the probability for the
quasimolecule to reach a certain internuclear distance $R_{\rm in}$ while still
remaining on the channel $e$. This can be obtained from calculations by 
monitoring
the quantum flux $J_e(R)$ on channel $e$ directly as is done when using the
OBE method, the MCWP method and the Landau-Zener approaches (two-state case).
Alternatively we can monitor the population transferred to the probe channel
$p$, as happens in the complex potential method (three-state case). 

We need a treatment that contains both the laser-induced excitation at
$R_C$ from $g$ to $e$, and survival on $e$. Furthermore, for strong fields a
proper model must allow any decayed population to be excited back to $e$ if
the decay has taken place in the vicinity of $R_C$. Only the MCWP method and
the OBE approach can handle this reexcitation (also called population
recycling) quantitatively. Further discussion on the subtle aspects of these
loss mechanisms can be found e.g.~in
Refs.~\cite{Smith,Suominen,experiments,JV}.

It should be pointed out that the OBE method, the MCWP method (as we apply it
in this paper), and the Landau-Zener approaches are ``one-way" studies. We only
consider the flux going in, but do not allow for any outcoming flux. For
estimating the FS and RE loss this is adequate as long as the detuning of the
laser is about 1-10 atomic linewidths. Then the role of bound states in the
loss mechanisms is not too important yet; see Ref.~\cite{JSB} for more
discussion. For strong fields the power broadening also diminishes the role
of the bound states; most of the loss is due to processes associated with the
first passage of the critical point $R_{\rm in}$ on the channel $e$.

We have earlier in a short article~\cite{Rapid} presented the results obtained
with the adiabatic OBE method and the MCWP method in the strong field
regime, and discussed the physical implications of the results (the lack of
saturation in trap loss when excitation becomes saturated). In this paper we
study in detail the various theoretical approaches, present the general
derivation of the OBE equations and their application to the two-state case
(Sec.~\ref{obe}) and show how to extend the Landau-Zener approach to the
strong field regime as suggested in Ref.~\cite{Suominen95} in the case of
optical shielding.   

The model Hamiltonian is
\begin{equation}
   H =	T(R)\mbox{\bf 1}+ \left(\begin{array}{cc}
       U_g(R)+U_c(R,l)&\hbar\Omega\\
       \hbar \Omega&    \hbar\Delta+U_e(R)+U_c(R,l)\end{array}\right),
       \label{hamiltonian1a}
\end{equation}
where $T(R)$ is the radial kinetic energy operator, $U_g(R)$ is the ground
electronic state which  behaves asymptotically as $U_g(R) = C_6/R^6$,
$U_c(R,l) = \hbar^2 l(l+1)/2\mu R^2$  is the centrifugal potential, $\mu$ is
the reduced mass of the quasimolecule (we assume a Cs$_2$ system), $U_e(R)$ 
is the excited state potential correlating asymptotically to $^2S_{1/2}$ +
$^2P_{3/2}$  alkali atoms and behaving asymptotically as $-C_3/R^3$,
and $\Delta$ is the detuning from resonance, $\hbar\Delta = E(P_{3/2}) -
E(S_{1/2}) - \hbar\omega$. Here $\omega$ is the laser frequency, and the
laser-induced coupling is described by the Rabi frequency $\Omega$.
The $g$ and $e$ channel potentials cross at the Condon point $R_C(\Delta)$
where $\hbar\Delta = U_e(R_C) - U_g(R_C)$. This crossing occurs at large
internuclear distances. The values used for potential parameters are
$C_3=20.30\ e^2a_0^2$ and $C_6=6.40 \times 10^5\ e^2a_0^2$, where $e$ is the 
electron charge and $a_0=0.529$ \AA\ is the Bohr radius. Note that we 
express all energy and angular frequency parameters in frequency units.

In studies using different laser parameters we have selected for a suitable 
inner distance $R_{\rm in}= 143\ a_0$, although the methods that we use allow 
the determination of $J_e(R)$ for all values of $R$. The atomic excited
state has a linewidth $\gamma_{\rm at}=5.13$ MHz, and we have taken the
molecular linewidth to be $\gamma=(4/3)\gamma_{\rm at}$, independent of $R$. 
And as mentioned before, we consider only the case $l=0$.

A reasonably complete description of cold atom collision dynamics in laser
traps can be obtained via the time-dependent density matrix $\rho(R,R';t)$
satisfying the Liouville equation~(\ref{liouville1}). The probability of
reaching the inner region on the excited state  potential (or the particle
flux in the inner region on the excited state potential) can be determined
directly from the diagonal density matrix elements $\rho_{ii}(R_{\rm in}, 
R_{\rm in};t)$. 
However, the direct numerical solution for the density matrix is presently
beyond our capabilities for the cold atom collision problem. Instead we shall
use various approximations to solve for the dynamics. The time-dependent
MCWP approach basically includes all the physics contained in the Liouville
equation, and it will provide the standard against which all other methods
are to be judged. 

Our model is a simplified representation of the true collision situation. 
Experiments have shown that inclusion of hyperfine  structure is necessary to
properly treat the collisions of laser  cooled alkali
species~\cite{experiments,Arimondo}.  
It is exactly for this reason that it is so 
important to develop simple and accurate approximate numerical methods that
can conceivably be used on problems including hyperfine dynamics where a
large number of channels is required to treat the manifold of the hyperfine
states.

\section{Optical Bloch equations}\label{obe}

\subsection{General situation}

In this section we derive the semiclassical optical Bloch equations. We start
with the Liouville equation~(\ref{liouville1}) and use Wigner function
description into which we introduce semiclassical approximations. If we write
Eq.~(\ref{liouville1}) by components we get
\begin{eqnarray}
   &&i\hbar\frac{\partial}{\partial t} \rho_{ij}(R,R';t) = -\frac{\hbar^2}
   {2\mu}\left(\frac{\partial^2}{\partial
   R^2}-\frac{\partial^2}{\partial R'^2}\right)
   \rho_{ij}(R,R';t) \nonumber\\
   && + \sum_k [V_{ik}(R)\rho_{kj}(R,R';t) -\rho_{ik}(R,R';t)
   V_{kj}(R')] + i\hbar \sum_{kl}\Gamma_{ijkl}\rho_{kl}(R,R';t). 
   \label{liouville2}
\end{eqnarray}
Here we have explicitly written out the kinetic and potential energy parts of
the Hamiltonian $H$; the term $V(R)$ contains the potentials for the  internal
states of the quasimolecule and couplings between them.  We assume that $V$
has no time-dependence as we have eliminated the oscillating laser field terms
using the rotating wave approximation and an appropriate phase shift (here
$V$ contains the potentials $U$ and couplings $\hbar\Omega$).

It should be pointed out that our description is time-dependent, so
instead of the boundary conditions used in the time-independent
scattering theory we have an initial value problem, i.e., we solve
Eq.~(\ref{liouville2}) starting at $t=t_0$ with some initial density matrix 
$\rho(R,R';t_0)$. If Eq.~(\ref{liouville2}) corresponds to a closed
system (no decay out of the selected set of levels), then $\rho$ can have 
steady state solutions. 

Typically one takes as the initial state the steady state result
corresponding to atoms being well separated, with
the quasimolecule potentials being flat over the distance that the system
moves within the time it takes to establish the steady state. In other
words, the molecular potentials do not impose any dynamics that would
interfere with the steady state formation. 

In time-independent scattering theory the initial conditions can not be stated
in terms of diabatic states if couplings between the states do not disappear 
asymptotically, as is the case with laser-induced quasimolecule processes. 
However, since the system at large $R$ evolves quickly into the local steady 
state, which is independent of
the selected basis states, there is no {\em a priori} reason to regard the 
adiabatic basis better than the diabatic basis. In practice one
tends to choose the diabatic basis, because it allows a simple description
of the spontaneous emission processes. Furthermore, as discussed later, we
can select {\em any} initial state in {\em any} basis, if we allow the system
initially enough time to reach locally at large $R$ a steady state before 
the spatial dependence of the quasimolecule potentials will couple the steady 
state formation and molecular dynamics. 

We assume for simplicity that in our current description $\Gamma$
is independent of position, but it is easy to extend our treatment to allow
$R$ dependence in $\Gamma$; such dependence can easily arise if retardation
effects are properly included to the quasimolecule potentials and lifetimes. We
redefine our spatial coordinate system by writing $R=r+q/2$, $R'=r-q/2$, which
transforms the kinetic term:
\begin{equation}
   \frac{\partial^2}{\partial R^2}-\frac{\partial^2}{\partial R'^2} =
   2\frac{\partial}{\partial r}\frac{\partial}{\partial q}.
\end{equation}

The density matrix $\rho(R,R';t)$ contains information about the spatial
coherences in the system. In order to calculate quantum fluxes at some
interatomic distance $R$ we do not need all that information, but only the
spatially diagonal elements $\rho(R,R;t)$. However, the evolution of these
diagonal elements depends on the off-diagonal $\rho(R,R';t)$ elements. By using
the Wigner function
\begin{equation}
   W_{ij}(p,r;t) = \int_{-\infty}^{\infty} dq \exp(-ipq/\hbar)
                    \rho_{ij}(r+\frac{1}{2}q,r-\frac{1}{2}q;t) 
                    \label{wigner1}
\end{equation}
we can include the spatial coherences and yet effectively work
with the spatially diagonal terms only.

First we apply the Fourier transform given in Eq.~(\ref{wigner1})  on both
sides of Eq.~(\ref{liouville2}) in order to obtain the equation of motion for
the Wigner function:
\begin{eqnarray}
   && \frac{\partial}{\partial t} W_{ij}(p,r;t)
      +\frac{p}{\mu}\frac{\partial}
      {\partial r} W_{ij}(p,r;t) =  \nonumber\\
   && \int_{-\infty}^{\infty} dq \exp(-ipq/\hbar) \left\{ \frac{1}{i\hbar}
      \sum_k [V_{ik}(r+q/2) \rho_{kj}(r+q/2,r-q/2;t)-\rho_{ik}(r+q/2,r-q/2;t)
      V_{kj}(r-q/2)] \right. \nonumber\\
   && + \left. \sum_{kl}\Gamma_{ijkl} \rho_{kl}(r+q/2,r-q/2;t)\right\}.
      \label{wigner-eom1}
\end{eqnarray}
One should note that above we have applied integration by parts in order to
replace $-i\hbar\frac{\partial}{\partial q}$ with $p$; this requires that
$\lim_{q\rightarrow\pm\infty} \rho(r+\frac{1}{2}q, r-\frac{1}{2}q;t)= 0$,
i.e., that the spatial coherences disappear as we move away from the
diagonal---this is a reasonable assumption.

If we integrate the Wigner function over momentum $p$ we get the spatial
probability distribution, which we can define as
\begin{equation}
   \tilde{\rho}_{ij}(r,t) =
   \frac{1}{2\pi\hbar}\int_{-\infty}^{\infty} dp
   W_{ij}(p,r;t) .\label{rhotilde1}
\end{equation}
The quantity $\tilde{\rho}_{ij}(r,t)$ equals $\rho_{ij}(r,r;t)$, as can readily
be seen by substituting the expression for $W_{ij}(p,r;t)$ in
Eq.~(\ref{wigner1}) into the right hand side of Eq.~(\ref{rhotilde1}) and
carrying out the integration over $p$.  So, by integrating
Eq.~(\ref{wigner-eom1}) over $p$ we get the equation of motion for
$\tilde{\rho}_{ij}(r,t)$:
\begin{equation}
   \frac{\partial}{\partial t}\tilde{\rho}_{ij}(r,t) +
   \frac{1}{\mu}\frac{\partial}{\partial r}
   \left[ \frac{1}{2\pi\hbar}\int_{-\infty}^{\infty} dp p
   W_{ij}(p,r;t)\right] =
   \frac{1}{i\hbar} \sum_k [V_{ik}(r)\tilde{\rho}_{kj}(r;t)
   -\tilde{\rho}_{ik}(r;t) V_{kj}(r)]
   +\sum_{kl}\Gamma_{ijkl}\tilde{\rho}_{kl}(r,t). \label{rhotilde-eom1}
\end{equation}
Here we have used the fact that there is no $p$-dependence on the right-hand
side of Eq.~(\ref{wigner-eom1}), so that if we perform the momentum
integration first, we obtain a $\delta(q)$ function, and thus the
integration over $q$ merely sets $q=0$. 

If a convenient method of evaluating the kinetic term were available, we could
use the result~(\ref{rhotilde-eom1}) to obtain exactly the diagonal elements of
$\tilde{\rho}$ at given $r$ and $t$, $\tilde{\rho}_{ii} (r,t)$, which is the
probability of being in channel $i$ at position $r$ and time $t$. However,
since we do not have any exact methods for calculating the second term in
Eq.~(\ref{rhotilde-eom1}), we shall estimate it using a WKB approach. The WKB
approximation for the density matrix element is given by
\begin{equation}
   \rho_{ij}(R,R';t)\simeq
   \frac{a_i(R;t)a^*_j(R';t)}{\sqrt{p_i(R)p_j(R')}}
   \exp\{i[\beta_i(R) - \beta_j(R')]/\hbar\} , \label{rhoWKB1}
\end{equation}  where $p_i(R)$ is the local classical momentum in state $i$,
$\beta_i(R)$ is the action at $R$,
\begin{equation}
   \beta_i(R) = \int^R dx\, p_i(x),\label{action1}
\end{equation}
and $a_i(R,t)$ is the amplitude factor for the WKB wave. We consider only the
incoming wave, which fixes the sign of the $\beta$ terms, and assume that
$a_i$ and $p_i$ do not depend on position very strongly. Furthermore, we
assume that the classical momenta $p_i(R)$ are non-zero and real. It should be
noted that by introducing the classical momenta we have made our equations
energy dependent as well, since
\begin{equation}
   p_i(r) = \sqrt{2\mu\{E-[V_{ii}(r)-V_{ii}(\infty)]\} },\label{classmom}
\end{equation} 
where $E$ is the asymptotic energy for the WKB wave (equal to the asymptotic
relative kinetic energy of the colliding atoms). 

It should be noted that our definition~(\ref{classmom}) of the classical 
momenta is clearly different from the one encountered in the traditional
scattering theory, if we consider the asymptotic situation. In the
time-independent theory the channels (states) are typically either open or 
closed, depending on the collision energy, i.e., their classical (WKB) momenta
are asymptotically either real or imaginary. This is because they are defined
as $p^{\rm scatt}_i(r) = \sqrt{2\mu\{E-[V_{ii}(r)-V_{00}(\infty)]\}}$, where
$i=0$ corresponds to the channel of the ingoing wave, defined by the
asymptotic boundary conditions. 

The difference here is due to the presence of the relaxation terms in 
Eq.~(\ref{liouville2}), and is required by the asymptotic situation. 
In the time-dependent treatment
we have $p_i(\infty) = \sqrt{2\mu E}$, which is independent of the state 
label $i$. This is because asymptotically we have a steady state formation
which is not coupled to the dynamics because the potentials are flat.
For simplicity we base our following discussion on a two-state system. Assuming
that the excited state and ground state populations had different asymptotic 
momenta, the steady state formation (the cycles of excitation and decay)
quickly mixes these populations and eventually the distribution of momentum on 
the ground state and the excited state would be exactly equal. In other words,
because of decay we can have asymptotic population even on a closed channel, 
but this population is a steady state reflection of the ground state 
population and must have the same classical momentum. Although in the above 
discussion we have assumed the diabatic basis, it is quite valid in the 
adiabatic basis, where the relaxation leads to a similar asymptotic mixing 
between the channels. Finally, as we fix the asymptotic situation by using 
Eq.~(\ref{liouville2}), we introduce other problems, which will be discussed 
in Sec.~\ref{implem}.

Next we insert Eq.~(\ref{rhoWKB1}) into Eq.~(\ref{wigner1}), and use the result
in Eq.~(\ref{rhotilde-eom1}). The exponential part of the integrand can be
expanded around $r$:
\begin{equation}
   \exp\{i[\beta_i(r+\frac{1}{2}q) - \beta_j(r-\frac{1}{2}q)]/\hbar\}
   \simeq \exp\{i[\beta_i(r) - \beta_j(r)]/\hbar
    +i\frac{1}{2}[p_i(r)+p_j(r)]q/\hbar + {\cal O}(r^2)\}.
\end{equation}
Then we apply the stationary phase method to obtain
\begin{equation}
   \frac{1}{2\pi} \int_{-\infty}^{\infty} dp p W_{ij}(p,r;t) = \frac{1}{2}
   [p_i(r) + p_j(r)] \tilde{\rho}_{ij}(r,t).   \label{int-wigner1}
\end{equation}
Substitution of this expression into Eq.~(\ref{rhotilde-eom1}) yields
\begin{equation}
   \frac{\partial}{\partial t}\tilde{\rho}_{ij}(r,t) +
   \frac{1}{2\mu} [p_i(r) + p_j(r)]
   \frac{\partial}{\partial r}\tilde{\rho}_{ij}(r,t) = \frac{1}{i\hbar}
   \sum_k [V_{ik}(r)\tilde{\rho}_{kj}(r;t) -\tilde{\rho}_{ik}(r;t) V_{kj}(r)]
   + \sum_{kl} \Gamma_{ijkl}\tilde{\rho}_{kl}(r,t) .\label{rhotilde-eom2}
\end{equation}
Here we have assumed that the classical momenta vary so little with $r$ that
they can be taken outside the derivative term.

Our aim is to find the total incoming quantum flux at each position $r$
(integrated over all times), and thus we are not interested in the actual time 
dependence. This simplifies our model to a great extent. Now the total flux 
can be obtained as a steady state result from Eq.~(\ref{rhotilde-eom2}). 
Since our model corresponds to a ``one-way" situation, the steady state flux
at $r$ is equal to the total flux that has passed that point. In the steady
state the time derivative in Eq.~(\ref{rhotilde-eom2}) vanishes, and we 
can replace $\tilde{\rho}_{ij}(r,t)$ by $\langle\tilde{\rho}_{kl} 
\rangle_{ss}(r)$. Furthermore, it is convenient to define the quantity 
\begin{equation}
   \sigma_{kl}(r) \equiv \sqrt{p_k(r)p_l(r)}\langle \tilde{\rho}_{kl}
   \rangle_{ss}(r) 
\end{equation}
whose diagonal elements give the flux in the various states at position $r$. 
We shall call this quantity the semiclassical density matrix. The equation of 
motion for it, within the validity range of the WKB approximation, is
\begin{equation}
   \frac{1}{2\mu} [p_i(r) + p_j(r)] \frac{d}{d r}\sigma_{ij}(r) =
   \sum_{kl}\left[
   \frac{p_i(r)p_j(r)}{p_k(r)p_l(r)}\right]^{1/2} [L_{ijkl}
   +\Gamma_{ijkl} ] \sigma_{kl}(r) ,	\label{rho-eom1}
\end{equation}
where
\begin{equation}
   L_{ijkl} = \frac{1}{i\hbar} [V_{ik}(r)\delta_{jl} -\delta_{ik}
   V_{lj}(r)].	\label{defineL}
\end{equation}
This completes the general derivation of the semiclassical optical Bloch
equations using the Wigner distribution. The set of 
equations~(\ref{rho-eom1}) were
obtained by making a semiclassical approximation, which focuses on classical
paths by virtue of using the WKB approximation. Only when the semiclassical
approximation is valid (when the de Broglie wavelength is smaller than the
region where the potentials are varying) will this approximation be meaningful.

The above derivation was done in the diabatic representation of the 
quasimolecule
potentials. The decay term $\Gamma$ has a simple form in this representation,
where the electronic states are also independent of position, and the internal
states are directly coupled by the standard dipole term. In the adiabatic
representation the electronic wavefunctions vary with the internuclear
coordinate $R$. In this presentation the laser-induced couplings and the decay
term become clearly $R$ dependent. We can move from the simple diabatic
representation into the adiabatic one, in which the potential matrix $V$
(which contains the radiative coupling) is diagonal. The transformation matrix
$C(R)$ is $R$ dependent,
\begin{equation}
   \sum_{kl} C_{ik}^{-1}(R) V_{kl}(R) C_{lj}(R) = \delta_{ij}E_{i}(R) ,
   \label{transformation1}
\end{equation}
where $E(R)$ is the diagonal eigenvalue matrix that gives the field-dressed
quasimolecule potentials. We can now write the diabatic semiclassical density
matrix in terms of the adiabatic semiclassical density matrix 
$\rho^a_{lk}(R,R';t)$:
\begin{equation}
   \rho_{ij}(R,R';t) = \sum_{kl}C_{ik}(R)\rho^a_{kl}(R,R';t) C^{-1}_{lj}(R').
   \label{transformation2}
\end{equation}
Next we insert this $\rho_{lk}(R,R',t)$ into Eq.~(\ref{liouville2}), and 
proceed as in the diabatic case. When making the
WKB approximation and using the stationary phase approach we assume that the
matrix elements $C_{ij}(R)$ are slowly varying functions in position. Thus, by
sandwiching the whole equation between $C^{-1}$ and $C$ we obtain eventually
\begin{equation}
   \frac{1}{2\mu} [p^a_i(r) + p^a_j(r)] \frac{d}{d r}\sigma^a_{ij}(r) =
   \frac{i}{\hbar} [E_j(r)-E_i(r)]\sigma^a_{ij} -\sum_{kl}
   \left\{Q_{ijkl}
   \frac{1}{2\mu} [p^a_k(r) + p^a_l(r)]-\Gamma^a_{ijkl}\right\}
   \sigma^a_{kl}(r)
   \left[ \frac{p^a_i(r)p^a_j(r)} {p^a_k(r)p^a_l(r)}\right]^{1/2},
   \label{rho-eom2}
\end{equation}
where the classical momenta $p^a_i(r)$ are now defined using the adiabatic
potentials $E_i(r)$ in Eq.~(\ref{classmom}). The decay term transforms as
\begin{equation}
   \Gamma^a_{ijrt} = \sum_{klmn} C_{ik}^{-1} C_{lj} \Gamma_{klmn} 
   C_{mr} C_{tn}^{-1}.\label{Gamma-ad}
\end{equation}
The non-adiabatic coupling that arose when we evaluated $\partial/\partial r
[C\tilde{\rho}^a(r,t)C^{-1}]$ is given by
\begin{eqnarray}
   Q_{ijkl}  &=& \sum_m \left[C^{-1}_{im}(r) \frac{\partial}{\partial r}
   C_{mk}(r)\delta_{jl} + \delta_{ik} \frac{\partial}{\partial r}
   C^{-1}_{jm}(r) C_{ml}(r)\right] \nonumber\\
   &=& \sum_m \left[C^{-1}_{im}(r) \frac{\partial}{\partial r} C_{mk}(r)
   \delta_{jl} - \delta_{ik} C^{-1}_{jm}(r) \frac{\partial}{\partial r}
   C_{ml}(r) \right] \nonumber\\
   &=& A_{ik}(r) \delta_{jl} - \delta_{ik} A_{lj}(r),
\label{defineQ}
\end{eqnarray}
where $A_{ik}(r) = \sum_m C^{-1}_{im}(r) \frac{\partial}{\partial r}
C_{mk}(r)$.

Another method for deriving the adiabatic OBE equations of motion involves 
using the half-collision matrix method~\cite{half}.  This method yields the 
same result as given by Eqs.~(\ref{rho-eom2}) and~(\ref{defineQ}) for the 
Hamiltonian part of the dynamics, but can not be used to derive the decay part
of the adiabatic OBE equations, since the half-collision method does not
incorporate the dissipative dynamics due to spontaneous emission contained 
in the density matrix treatment.

\subsection{The two-state case}

\begin{figure}
\centerline{\psfig{width=130mm,file=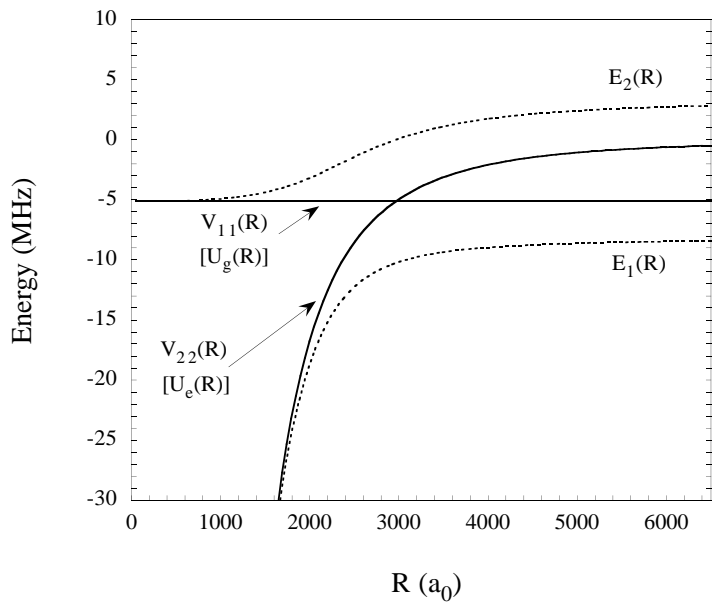}}
\caption[f2]{The two-state model for trap loss collisions. The solid lines
describe the bare (diabatic) quasimolecule potentials near the Condon point
$R_C$. Here $\Delta=\gamma$, so $R_C\simeq 2930$ a$_0$. The dotted lines
describe the field-dressed (adiabatic) potentials.\label{potsut}}
\end{figure}

We assume that our quasimolecule has only two states, one ground state
(1) and one excited state (2), with potentials $V_{11}$ and $V_{22}$. In our
trap loss model these states are as shown in Fig.~\ref{potsut}. The excited
state has a constant width $\gamma$, and the off-diagonal density matrix
elements ($\rho_{12}$ and $\rho_{21}$) have the width $\frac{1}{2}\gamma$. 
Thus we have
\begin{equation}
   \Gamma_{1122}=\gamma,\quad \Gamma_{2222} = -\gamma,\quad
   \Gamma_{2121}=-\frac{1}{2}\gamma,\quad\Gamma_{1212}
   =-\frac{1}{2}\gamma,
\end{equation}
and rest of the elements of $\Gamma$ are zero. Using $V_{12}$ as
the coupling between the states, we get the diabatic equations
\begin{eqnarray}
   \frac{d\sigma_{11}}{dr} &=& \frac{i}{\hbar} \frac{V_{12}}{\sqrt{v_1
      v_2}}(\sigma_{12}-\sigma_{21}) +\gamma \frac{\sigma_{22}}{v_2}
      \label{1.6a}\\
   \frac{d\sigma_{22}}{dr} &=& -\frac{d\sigma_{11}}{dr} \label{1.6b}\\
   \frac{d\sigma_{12}}{dr} &=& \frac{i}{\hbar}\frac{2(V_{11}-V_{22})+i\gamma}
      {v_1+v_2}\sigma_{12}-\frac{i}{\hbar}\frac{2V_{12}}{v_1+v_2}
      \left( \sqrt{\frac{v_1}{v_2}} \sigma_{22} - \sqrt{\frac{v_2}{v_1}}
      \sigma_{11} \right) \label{1.6c}\\
   \frac{d\sigma_{21}}{dr} &=& \frac{d\sigma_{12}^*}{dr}  . \label{1.6d}
\end{eqnarray}
Here we have used the classical velocities $v_i(r)=p_i(r)/\mu$.

In order to move into the adiabatic frame we need the transformation matrix
elements, $C_{ij}(r)$. In a two-state system it is convenient to define
\begin{equation}
   \theta =\frac{1}{2} \tan^{-1}\frac{2V_{12}}{V_{22} - V_{11}} . \label{1.10}
\end{equation}
The transformation matrix is given by
\begin{equation}
   C(r) =  \left ( \begin{array}{rr} C_{11} &C_{12} \\
           C_{21} & C_{22} \end{array}\right )
        =  \left ( \begin{array}{rr} \cos(\theta) &\sin(\theta) \\
          -\sin(\theta) & \cos(\theta) \end{array}\right ),
\end{equation}
and the inverse transformation is obtained from the relation
$C^{-1}(\theta)  = C(-\theta)$. Thus we get
\begin{equation}
   A = C^{-1}\frac{\partial C}{\partial r} = \frac{\partial \theta}{\partial r}
       C^{-1}\frac{\partial C}{\partial \theta}
     = \frac{\partial \theta}{\partial r} \left ( \begin{array}{rr} 0 & 1 \\
         -1 & 0 \end{array}\right ).
\end{equation}
For convenience, we define $D=\frac{\partial \theta}{\partial r}$. This gives
us
\begin{equation}
   Q_{1121}=Q_{1222}= A_{12}= D,\quad Q_{1112}=-A_{21}=D,\quad
   Q_{1211}= -A_{12} = -D.
\end{equation}

After a little algebra, we get the semiclassical optical Bloch equations
in the adiabatic representation:
\begin{eqnarray}
   \frac{d\sigma^a_{11}}{dr} &=& -D\frac{v^a_1+v^a_2}{2\sqrt{v^a_1v^a_2}}
                                 (\sigma^a_{21} +\sigma^a_{12})
                                 -\gamma \left[ \frac{s^4}{v^a_1}
                                 \sigma^a_{11} -\frac{c^4}{v^a_2} \sigma^a_{22}
                                 +\frac{sc(c^2-s^2)}{2\sqrt{v^a_1v^a_2}}
                                 (\sigma^a_{21}+\sigma^a_{12})
                                 \right] \label{1.7a}  \\
   \frac{d\sigma^a_{22}}{dr} &=& - \frac{d\sigma^a_{11}}{dr} \label{1.7b}\\
   \frac{d\sigma^a_{12}}{dr} &=& - \frac{2i}{\hbar} \frac{E_{1}-
                                 E_{2}}{v^a_1+v^a_2} \sigma^a_{12} -
                                 D\frac{2\sqrt{v^a_1v^a_2}} {v^a_1+v^a_2}
                                 (\sigma^a_{22}-\sigma^a_{11})\\
                             & & -\frac{\gamma}{v^a_1+v^a_2} \left\{
                                 \sigma^a_{12} + 2 s^2 c^2 (\sigma^a_{21}
                                 +\sigma^a_{12}) - \left[ sc(1+2 c^2)
                                 \sqrt{\frac{v^a_1}{v^a_2}}
                                 \sigma^a_{22} + sc(1+2 s^2)
                                 \sqrt{\frac{v^a_2}{v^a_1}} \sigma^a_{11}
                                 \right]\right\}  \label{1.7c}\nonumber\\
   \frac{d\sigma^a_{21}}{dr} &=& \frac{d{\sigma^a_{12}}^*}{dr}.
                                 \label{1.7d}
\end{eqnarray}
Here we use the notation $s=\sin(\theta)$ and $c=\cos(\theta)$.
The velocity factors $v^a_1$ and $v^a_2$ are defined as in 
Eq.~(\ref{classmom}), using $E_1$ and $E_2$ as the appropriate potentials.
At this point we relabel $r$ with $R$.

\subsection{The implementation of the two-state case}\label{implem}

The optical Bloch equations (\ref{1.6a})-(\ref{1.6d}) and
(\ref{1.7a})-(\ref{1.7d}) can be solved numerically using various methods. As
will be discussed in Sec.~\ref{results}, the diabatic formulation fails when
$T< 1$ mK for all coupling strengths $\Omega$. If we use the adiabatic
equations~(\ref{1.7a})-(\ref{1.7d}), with the velocity factors given by
Eq.~(\ref{classmom}), we find a good agreement with the MCWP results for all
temperatures at large couplings, but an increasing deviation with decreasing
$\Omega$. Of course, we would like to have an approach which is good for
all $T$ and $\Omega$. Thus we need to look into the issue of velocity factors
in detail.

The velocity factor $v_2^a$ for the upper adiabatic potential (see
Fig.~\ref{potsut}) is the source of the problem here. When the atoms approach 
each other, a steady state is quickly formed and maintained until the system 
gets very close to $R_C$. As discussed earlier, this gives us the advantage 
of a basis-independent initial state. At this point the population on the 
upper adiabatic state 2 is a steady state reflection of the population of the 
lower adiabatic state 1. Thus when the system reaches the Condon 
point the probability flux on both adiabatic states ought in fact have the 
same momentum. Thus, if we use the velocity factor given by 
Eq.~(\ref{classmom}) for the upper adiabatic state, we get the right velocity 
asymptotically, but the wrong velocity at the crossing. We can correct this 
by redefining $v_2^a(R)$ so that $v_2^a = \sqrt{2E/\mu}$ for all $R$. 

In other words, because the OBE approach mixes concepts from time-dependent
theory (steady state formation) and time-independent theory (WKB wave
functions), we must sometimes improve its performance by such a tuning. 
Here we need to give the WKB wave functions asymptotically the velocities 
which describe the steady state situation correctly, but at the crossing
the dynamics dominates, implying that the classical momenta
used in the time-independent theory would provide a more accurate description.
Deviations are visible in the limit of the weak excitation, because then the 
excited state population is dominated by the small steady state contribution 
which, due to the large momentum given incorrectly by Eq.~(\ref{classmom}), 
has a good chance to survive on the excited state until $R=R_{\rm in}$. This
population overwhelms the contribution from the dynamical excitation, and
gives thus incorrect results. However, as will be shown in Sec.~\ref{results}, 
we now have a method that can predict correctly the probability to reach small 
$R$ on channel $e$ for any practical $\Omega$, and for the temperatures in the 
cold collisions regime. It should be noted that an extension of our approach 
to ultracold temperature regime, i.e., below the recoil limit, is not likely 
to succeed, as the semiclassical viewpoint fails in this 
regime~\cite{BECPRL,JNIST}. 

The previous formulation of optical Bloch equations~\cite{BJ} was done in
the time frame rather than in the position frame. The transition between 
these two frames was performed by introducing a reference trajectory $v_0(R)$, 
which mapped $R$ to $t$. However, as noted in Ref. [14], this reference 
trajectory is not really needed as we can do the calculation in the $R$ 
frame altogether. 

In practice when solving the optical Bloch equations we can set all
$\sigma_{ij}(R_0)=0$, except selecting one state for which
$\sigma_{ii}(R_0)=1$. The value of $R_0$ is set suitably large for the atoms
to be well apart and potentials flat. Since the classical trajectory couples 
$R$ and $t$, evolution in $R$ corresponds to evolution in $t$ and we find that
the system has evolved into a steady state distribution of the ground and 
excited state populations after moving a relatively short distance towards 
smaller $R$. This corresponds to numerically determining the asymptotic state 
state populations and coherences.

\section{Monte-Carlo simulations}\label{MC}

A direct wave packet treatment of Eq.~(\ref{liouville1}) with numerical
methods is possible \cite{Lai93,SG93}. However, since one has to
operate with a two-dimensional spatial grid, the memory sizes currently
available in computers strongly limit the use of this approach: only models
which are simplified in the extreme can  be studied.  In cold collisions the
acceleration of the initially slow  wave packet on the steep excited state
potential surface forces us to  use a large two-dimensional momentum space
while at the same time good  momentum resolution is needed to define
adequately the narrow initial wave packet for low temperatures.  Similar
demands are set for the position space as well; for more detailed discussion
see  Refs.~\cite{SG93,HSB94}.  Hence we need many grid points in order
to span properly the required regions in both the momentum and position
spaces.

It may be feasible to avoid some of the computational problems by using
grid sizes and resolutions which are adaptive; one might utilize the
rather deterministic behavior of the wave packet by altering the
computational grid properties either as a function of position or time. 
Then the straightforward swapping between momentum and position
representations using fast Fourier transforms is, however, usually lost.  We
have chosen to approach the problem from another angle.  The Monte Carlo wave
packet (MCWP) method allows us to treat Eq.~(\ref{liouville1}) numerically
as a one-dimensional problem.  Unlike other approximative methods this 
approach does not adapt any concepts from classical mechanics and, therefore,
it is not a semiclassical tool but a fully quantum one. Hence it can be used
as a benchmark for the different semiclassical methods described in this
article.

By using the MCWP method we can greatly diminish the limitations set by the 
available computer memory. However, this gain is partially reduced by the
increase in the time required by the computation. Hence the MCWP simulations
are quite time-consuming, and therefore the need for other approaches is
quite acute. There are several Monte-Carlo approaches available currently,
and we use the Dalibard-Castin-M{\o}lmer version~\cite{DCM92}, adapted to wave
packet problems in the manner described in
Refs.~\cite{Lai93,HSB94,Suominen95}.  We shall give here a brief description
of the method, but keep the main emphasis on aspects related to the 
particular system studied in this article.

In the MCWP method one does not directly solve the time evolution of the
density matrix itself. Instead, we look at the time evolution of the state
vector
\begin{equation}
  \Psi(R,t)=\left(\begin{array}{c} \Psi_g(R,t) \\
             \Psi_e(R,t)\end{array}\right), \label{statevector}
\end{equation}
where $\Psi_g$ and $\Psi_e$ are the ground and excited state probability 
amplitudes, respectively. In this model the spontaneous decay appears as 
random quantum jumps during the time evolution of the state vector. Hence
each time we solve the time-dependent Schr\"odinger equation we obtain a
unique state vector evolution, also called a wave packet history. One can
form a finite ensemble of such histories and calculate ensemble averaged
expectation values for physical quantities. These values are approximations
to those provided by the full density matrix treatment. The accuracy of
the ensemble averages tends to increase with the number of members in the
ensemble, and in the limit of an infinite ensemble these averages and the
density matrix results become equal, as shown e.g.~in Ref.~\cite{DCM92}. So,
we expect that by accumulating ensemble members we can eventually reach a
suitable accuracy at some finite ensemble size. The accuracy to be expected of
the method is decribed in  Ref.~\cite{DCM92}. For wave packets in cold
collision problems the appropriate ensemble size seems to be roughly 50
members, assuming that all the histories  are very close to each other in the
phase space for all times. This is quite true for the excited state survival
studies related to attractive excited states. In general such localisation in
the phase space for all times is necessary for the success of the
semiclassical approaches.

We start the calculation of the state vector evolution from some initial
state, which in our case is a Gaussian wave packet on the ground state moving
towards small $R$. The wave packet in general describes the probability to
find the two colliding atoms at certain relative separation $R$, and its
components $|\Psi_g(R,t)|^2$ and $|\Psi_e(R,t)|^2$ contain the additional
information how this  probability is distributed between the ground and
excited states.  A Fourier transform of the state vector $\Psi(R,t)$ takes
the system into the momentum representation. We set the initial phase of 
$\Psi_g$ in position representation such that the wave packet starts with a
mean momentum $\langle p\rangle$ which corresponds to the temperature of the
cloud of cooled and trapped atoms.  The width of the wave packet is chosen so
that it remains relatively narrow in both  representations.  We cannot, of
course, violate the Heisenberg uncertainty relation, so it is impossible to
have infinitely narrow  packets in either representation. It should be
pointed out that apart from satisfying the Heisenberg uncertainty relation the
width of the wave  packet is not related to any of the macroscopic quantities
of the  physical situation which we try to simulate.

Initially the wave packet is located far from the crossing so that a steady
state between the ground and excited states can form before the wave packet
reaches the interaction region where the dipole-dipole interaction makes the
laser resonant with the molecular transition. The time scale for the
formation of the steady state is roughly 3-5 times the decay time scale
$1/\gamma_{\rm mol}$ \cite{Loudon}, assuming that the local detuning does not
change much over the distance covered by the wave packet during that time.  
As discussed before, the steady state formation allows us to put the initial 
wave packet on the diabatic ground state even when the laser-induced
coupling is large asymptotically, because the final steady state is 
independent of the initial state. Indeed, we could even place the initial 
wave packet on the excited state, and yet the wave packet approaching the 
crossing would still be the steady state one.

The state vector corresponding to the initial state is stepped forward 
in time with its evolution determined by the Schr\"odinger equation.  Various
numerical methods can be used, and we have applied the combination of split
operator approach with  Crank-Nicholson and Runge-Kutta algorithms, described
in detail in Ref.~\cite{HSB94}. In the MCWP method one uses an effective 
Hamiltonian,
\begin{equation}
   H_{\rm eff} = H-i\frac{\hbar\gamma}{2} \sigma^+\sigma^-,
\end{equation}
where $H$ is the system Hamiltonian, $\gamma$ is the decay rate for the
excited state population and $\sigma^+$ and $\sigma^-$ are the standard 
spinor raising and lowering operators, respectively.

For each time step $t\rightarrow t+\delta t$ we calculate the jump probability
\begin{equation}
   \delta s = \gamma P_e(t)\delta t,
\end{equation}
where $P_e(t)$ is the current excited state population.  By rewriting $\delta
s$ as $dP_e$ we would end up with the standard exponential decay $\exp(
-\gamma t)$ of the excited state population.  Now, we  continue by comparing
the jump probability $\delta s$ with a random number $\eta\in [0,1]$. A
quantum jump occurs when $\eta<\delta s$; this is usually the less likely
situation since the  basic assumption in the derivation of the MCWP method is
that $\delta  s\ll 1$ all the time (guaranteed by choosing $\delta t\ll
1/\gamma$).   When a jump occurs one simply replaces $\Psi_g(R,t+\delta t)$
with $\Psi_e(R,t+\delta t)$, and then sets $\Psi_e(R,t+\delta t)=0$.  The 
occurrence of the jump corresponds to the observation of a fluorecence
photon, which reduces the wave function: before the jump it had to be in the
excited state, and after the jump it must be in the ground state.  The
important aspect is that as the jump takes place the  position and momentum
properties of the excited state component of the state vector are transfered
to the ground state component.  This is the source of radiative heating,
among other things.

Both the evolution under $H_{\rm eff}$ and the quantum jumps reduce the norm
of the state vector $\Psi$.  Hence after each time step the  state vector is
renormalized to unity, even if a jump does not occur.  It should be noted
that if we had $H_{\rm eff}=H$, then we would always observe a quantum jump
for a system with non-zero $P_e$ if we wait long enough.  For cases where
$P_e<1$ this would be wrong, since there is a non-zero probability that the
system never was on the excited state.  In the weak field limit $P_e$ is
always very small, so most of the ensemble members correspond to the time
evolution under the non-Hermitian Hamiltonian with no interruption by jumps. 
Then the wave packet approach reduces to a time-dependent version of the 
complex potential approach.  Therefore a single ensemble member becomes a
reasonably accurate approximation to the density matrix result.  We have used
this property in our weak field study~\cite{Suominen94}, and have thus
verified that the diabatic formulation of the OBE method does not work
properly at low temperatures, but the Landau-Zener approach and the complex 
potential method can be used instead.

Although the MCWP method allows us to use relatively large grids, the strong
change in the excited state potential corresponds at our probing distance
$R_{\rm in}=143\ a_0$ to kinetic energies which are beyond the numerical 
treatment. Basically it becomes impossible to correctly track the relevant 
quantum mechanical phase term $\exp(-i E\delta t/\hbar)$, where $E$ is the 
kinetic energy of the wave packet. Hence we must cut the excited state 
potential change by making it flat for $R_{\rm in}<R<R_{\rm cut}$; in our 
studies we have used the value $R_{\rm cut} = 512\ a_0$. At $R_{\rm cut}$ 
we are basically left with the exponential decay of the excited state 
population because of the large local detuning. Hence we can take the wave 
packet result for $J_e(R)$ from $R = R_{\rm cut}$ to $R_{\rm in}$ by 
multiplying it with $\exp(-\gamma t_{\rm  trans})$, where $t_{\rm trans}$ is 
the time it takes to go from $R_{\rm cut}$ to $R_{\rm in}$ along the classical
path determined by the local velocity. In fact, the same approach is also 
applied when the OBE equations are solved numerically: otherwise the adequate 
determination of the term $\exp[-i p(R) \delta R/\hbar]$ would require 
unpractically small values of the spatial grid spacing $\delta R$
(here $p(R)$ is the local momentum at $R$) when $R<R_{\rm cut}$.

\section{Complex potential calculations}\label{complex}

In the complex potential method one adds a complex term on the excited state
potential in order to describe decay out of these states~\cite{JSB,TB,BVT}.
This approach does not allow any reexcitation, and is thus not appropriate for
strong field studies directly (by including the photon states explicitly one
might improve the model although this would drastically increase the number of
channels required to solve even the case of two quasimolecule
states~\cite{BVT}).

In this method one simply uses the Hamiltonian
\begin{equation}
   H =	T(R)+ \left(\begin{array}{ccc}
       U_g(R)+U_c(R,l)&\hbar\Omega& 0\\
       \hbar \Omega&    \hbar\Delta+U_e(R)+U_c(R,l)-i\hbar\gamma/2&
       \hbar\Omega_{ep}\\
       0& \hbar \Omega_{ep}&E_p+U_p(r)+U_c(R,l)\end{array}\right),
       \label{cplxshamil}
\end{equation}
and solves the time-independent Schr\"odinger equation
\begin{equation}
  	\frac{d^2F(R)}{dR^2} + \frac{2\mu}{\hbar^2}[E -U(R)] F(R) = 0,
 		\label{schrod}
\end{equation}
where $F$ is the three-component state vector for our model and $E$ is the
asymptotic collision energy. 

For this method we have explicitly included the probe channel $p$ to our 
Hamiltonian:  $U_p(R)$ is the corresponding potential. The probe channel $p$ 
crosses the $e$ channel potential at $R_{\rm in}$. Because of the disparity 
between $R_C(\Delta)$ and $R_{\rm in}$ the outer zone excitation process is 
in practice well separated from the inner zone process. The coupling 
$\Omega_{ep}$ depends on the nature of the coupling of the excited state and 
the probe channel. 

We have used the invariant imbedding method~\cite{TB,invimb} (in the
diabatic representation) to solve the above close coupling equations in a form
that directly computes the $S$ matrix elements, $S_{gp}$ and $S_{ep}$, from
which the quantum flux $J_e(R)$, i.e., the quantum mechanical probability of
reaching the inner zone, is determined. In the complex potential calculations 
we use mainly the same values for parameters as in the two-state model, given 
in Sec.~\ref{model}. However, here the excited state potential, $U_e(R)$,
is taken as a numerical spline having a minimum energy of $-182$ GHz at $R =
72\ a_0$ and an asymptotic behavior of $-C_3/R^3$. We take the probe potential 
to be $U_p(R) = C_3^p/R^3$, with $C_3^p = 7.260\ e^2a_0^2$.  All potentials 
have repulsive inner walls so the $R<0$ region is non-classical.  The other 
parameters are $E_p=-3.0$ GHz and $\Omega_{ep}= 1.0$ MHz. 

As we are now working with the time-independent scattering theory, we need to
consider boundary conditions instead of initial ones. Since the model 
potential contains non-vanishing off-diagonal elements, for
strong fields there is a clear mixing of states at large $R$.  Therefore 
the boundary condition for the complex potential approach has to be defined in
terms of the field-dressed, i.e., adiabatic states. The transformation
between the different bases then takes place as described in Sec.~\ref{obe}. 
We assume an incoming wave (corresponding to the asymptotic collision energy 
$E$) on the lower adiabatic channel formed by states $g$ and $e$
(state 1 in Fig.~\ref{potsut}). 

\section{Landau-Zener approaches}\label{LZ}

Since the inner and the outer crossings are isolated in distinctly different
regions, we can write the loss probability, i.e., the probability to exit on
channel $p$ as
\begin{equation}
   P_p = |S_{gp}|^2 = P_{ep}(R_{\rm in}) J_e(R_{\rm in}),
\end{equation}
where the probability $P_{ep} = |S_{ep}|^2$ measures the quantum probability
of the inner zone $e\rightarrow p$ process due to traversing the inner curve
crossing once in both directions. Here $R_{\rm in}$ is the location of this
curve crossing. When using the complex potential method we obtain the flux
simply by using the relation $J_e(R_{\rm in}) = |S_{gp}|^2 / P_{ep}(R_{\rm
in})$. The probability $P_{ep}$ is almost completely insensitive to the laser
intensity $I$ and to the collision energy $E$ for small detunings.

In the limit of large detuning and small laser intensity it is safe to assume
that the excitation becomes localized to the Condon point $R_C$. In this
limit the dynamical models and the local equilibrium model tends to agree. In
local equilibrium models one assumes that the motion of the atoms is very slow
compared to the steady state formation, and thus the steady state formation 
dominates~\cite{GP}. This leads to a picture where off-resonant excitation is 
important. One can express the local excitation in a 
two-state model in the steady state regime as~\cite{Loudon}
\begin{equation}
   \sigma_{ee} = \frac{\Omega^2}
   {\Delta(R)^2+2\Omega^2+(\gamma/2)^2}\label{steady},
\end{equation}
where $\Delta(R)$ is the local detuning, $\hbar\Delta(R) = U_e(R)-U_g(R)
+\hbar\Delta$. As a final stage in the local equilibrium model one weights
the results with Eq.~(\ref{steady}) and integrates over the
position coordinate $R$.

In the dynamical models it is assumed that as the system approaches $R_C$ the
motion and thus the change in the local detuning become fast compared to the
steady state formation, and thus the excitation becomes a dynamical process
which is localized to a region near $R_C$. The dynamical excitation can then
be described with the Landau-Zener curve crossing
model~\cite{JSB,HSB94,Suominen94,Suominen95}. In steady state models for large
detunings the integration over the linewidth function (\ref{steady}) becomes
like a $\delta$ function which singles out the Condon point, and thus the two
viewpoints agree in this limit. The MCWP simulations have so far supported the
dynamical picture over the local equilibrium picture; for further discussion
see~Ref.~\cite{Suominen}.

In the weak field limit we can assume that the excitation and subsequent
decay are uncoupled, and reexcitation is negligible. Then we can write, using
the Landau-Zener model, the expression for the flux on channel $e$ as 
\begin{equation}
   J^{\rm LZD}_e(R) = S_e(R,R_C;E,\gamma)P_{\rm LZ}=
   S_e(R,R_C;E,\gamma)[1-\exp(-2\pi\Lambda)] .	\label{kolkytyks}
\end{equation} 
Here
\begin{equation}
   \Lambda = \frac{\hbar\Omega^2}{\alpha v_g(R_C;E)} ,	
   \label{kolkytkaks}
\end{equation} 
where $v_g(R;E)$ is the classical velocity associated with the ground state at
position $R$ when the collision energy is $E$, and
\begin{equation}
   \alpha = \left|\frac{dU_e(R)}{dR} -\frac{dU_g(R)}{dR}\right|_{R=R_C}.
\end{equation}
In other words, $P_{\rm LZ}$ is the one-way Landau-Zener probability of 
undergoing a transition from channel $g$ to channel $e$ at the Condon point
$R_C(\Delta)$, and $S_e(R,R_C;E,\gamma)$ is the survival probability
from $R_C$ to $R$. We can calculate the survival probability by assuming a
classical trajectory combined with exponential (Weisskopf-Wigner) decay:
\begin{equation}
   S =\exp(-\gamma t_{\rm cl});\qquad t_{\rm cl}(R,R_C;E)=\int^R_{R_C} 
   \frac{dR}{v_e(R)},\label{surv1}
\end{equation}
where $v_e(R)$ is the classical trajectory velocity for the excited state (in
the diabatic formulation). This is the Landau-Zener model with decay (LZD).

The above model fails when excitation and decay do not decouple, which happens
at strong fields due to reexcitation of decayed
population~\cite{HSB94,Suominen95}. We can think of reexcitation as a process
which delays the start of the exponential decay. Reexcitation takes place
mainly within some region around the Condon point. We can define an interaction
region for which $\Delta(R)<\Omega$. By making the simple assumption that
exponential decay can take place only outside this region, we can rewrite the
$t_{\rm cl}$ in Eq.~(\ref{surv1}) as
\begin{equation}
   t_{\rm cl} =  \int^R_{R_{\Omega}} \frac{dR}{v_e(R)},
\end{equation}
where $R_{\Omega}$ is defined by the relation $\Delta(R_{\Omega})=\Omega$;
$R_{\Omega}<R_C$. The modified survival term depends now on the laser
intensity $I$ ($I\propto\Omega^2$). We call this approach the Landau-Zener
model with delayed decay (LZDD). Obviously the model can only give qualitative
predictions, especially as the concept of the edge of the interaction region
is not well defined. However, it gives a good intuitive undestanding why
$J_e(R)$ for small $R$ keeps increasing even when the excitation saturates to
unity (and thus the LZD prediction saturates)~\cite{Rapid}. This picture
agrees qualitatively with the results from the OBE and MCWP calculations, as
will be shown in the next section.

\section{Comparison of methods}\label{results}

\begin{figure}
\centerline{\psfig{width=160mm,file=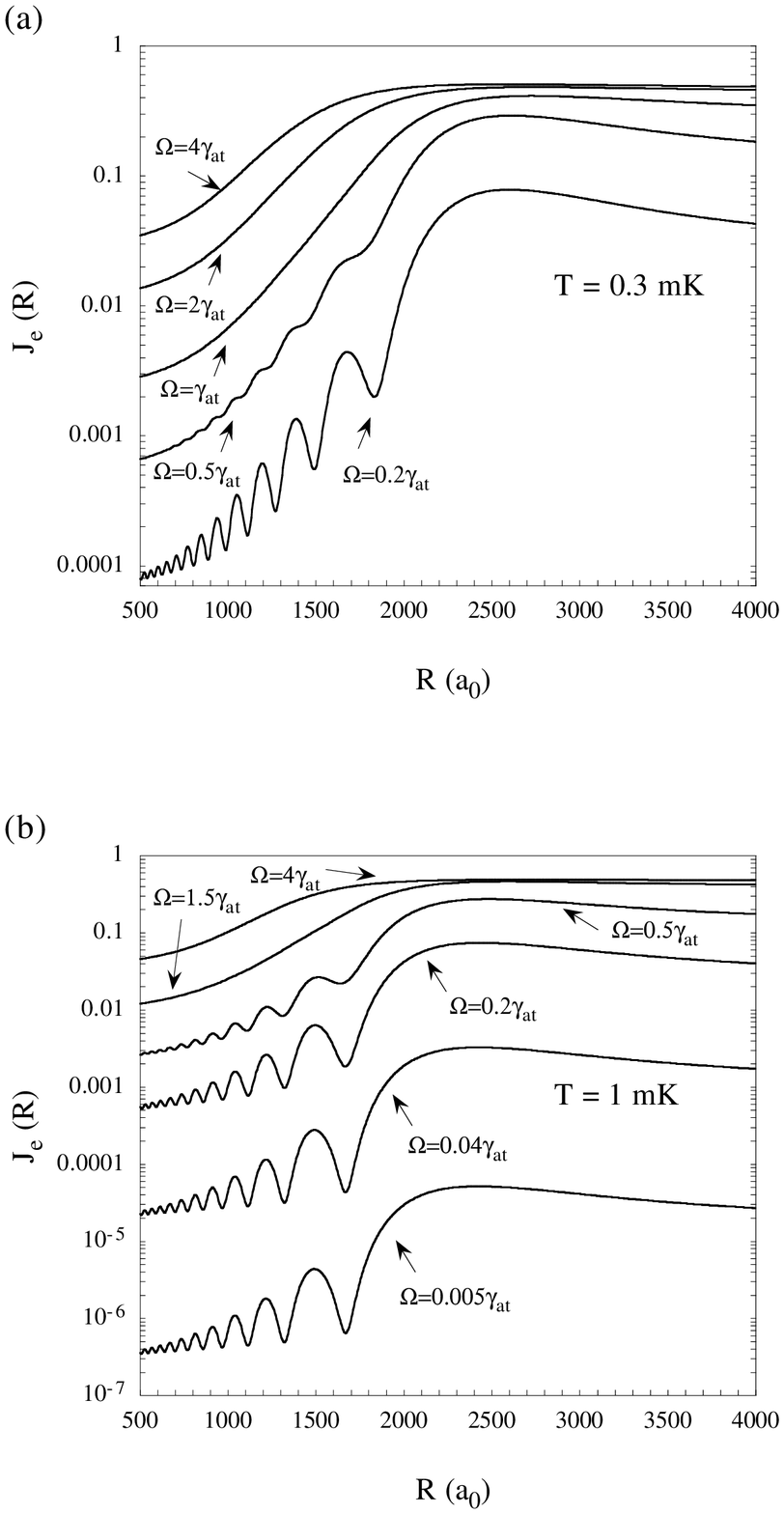}}
\caption[f3]{The excited state flux $J_e(R)$ calculated using the MCWP method.
Here $\Delta=\gamma_{\rm at}$, and the other parameters are as indicated in the
figure.\label{fluxR}}
\end{figure}

Typically the excited state flux $J_e(R)$ shows oscillations at $R<R_C$ in the
bare state picture. This is demonstrated by the MCWP results given in
Fig.~\ref{fluxR}. The oscillations are due to the coherences between the
two states, established near the Condon point. 
As the coupling $\Omega$ increases, 
the situation becomes increasingly adiabatic and the oscillations disappear. At
the same time the asymptotic (large $R$) flux approaches the steady state
result, 1/2.  It is interesting to note that although the main change in the
flux seems to take place over a wide region in $R$, the dynamical view with
excitation localized to $R_C$ works well, as demonstrated by us earlier in
Ref.~\cite{Suominen94}. 

\begin{figure}
\centerline{\psfig{width=160mm,file=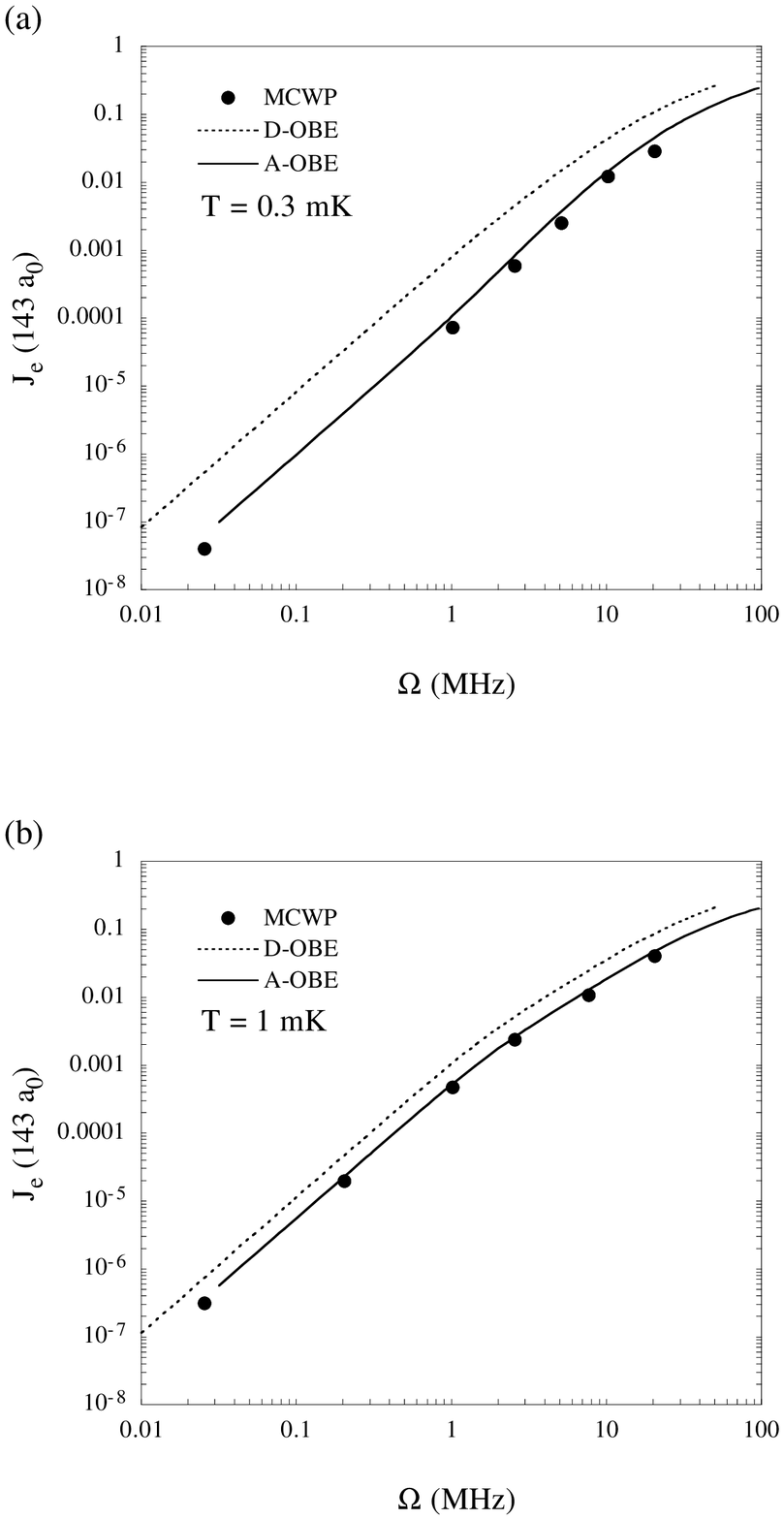}}
\caption[f4]{The excited state flux $J_e$ at $R=143\ a_0$ as a function of the
laser-induced coupling $\Omega$, calculated using the MCWP, D-OBE and A-OBE
methods. Here $\Delta=\gamma_{\rm at}$, and the other parameters are as
indicated in the figure.\label{MCWP-OBE}}
\end{figure}

In Fig.~\ref{MCWP-OBE} we show a comparison between the MCWP results, the
diabatic OBE results (D-OBE) and adiabatic OBE results (A-OBE). The agreement
between the MCWP and A-OBE results is very satisfactory, whereas the D-OBE
results fail by an order of magnitude for $T=0.3$ mK. This failure increases
further as $T$ decreases, as shown in Refs.~\cite{JSB,Suominen94}. As
discussed in the previous paragraph, the difference between D-OBE and
A-OBE results suggest strongly that the basic condition for the validity
of the local equilibrium model is not fulfilled for typical trap parameters. 

\begin{figure}
\centerline{\psfig{width=160mm,file=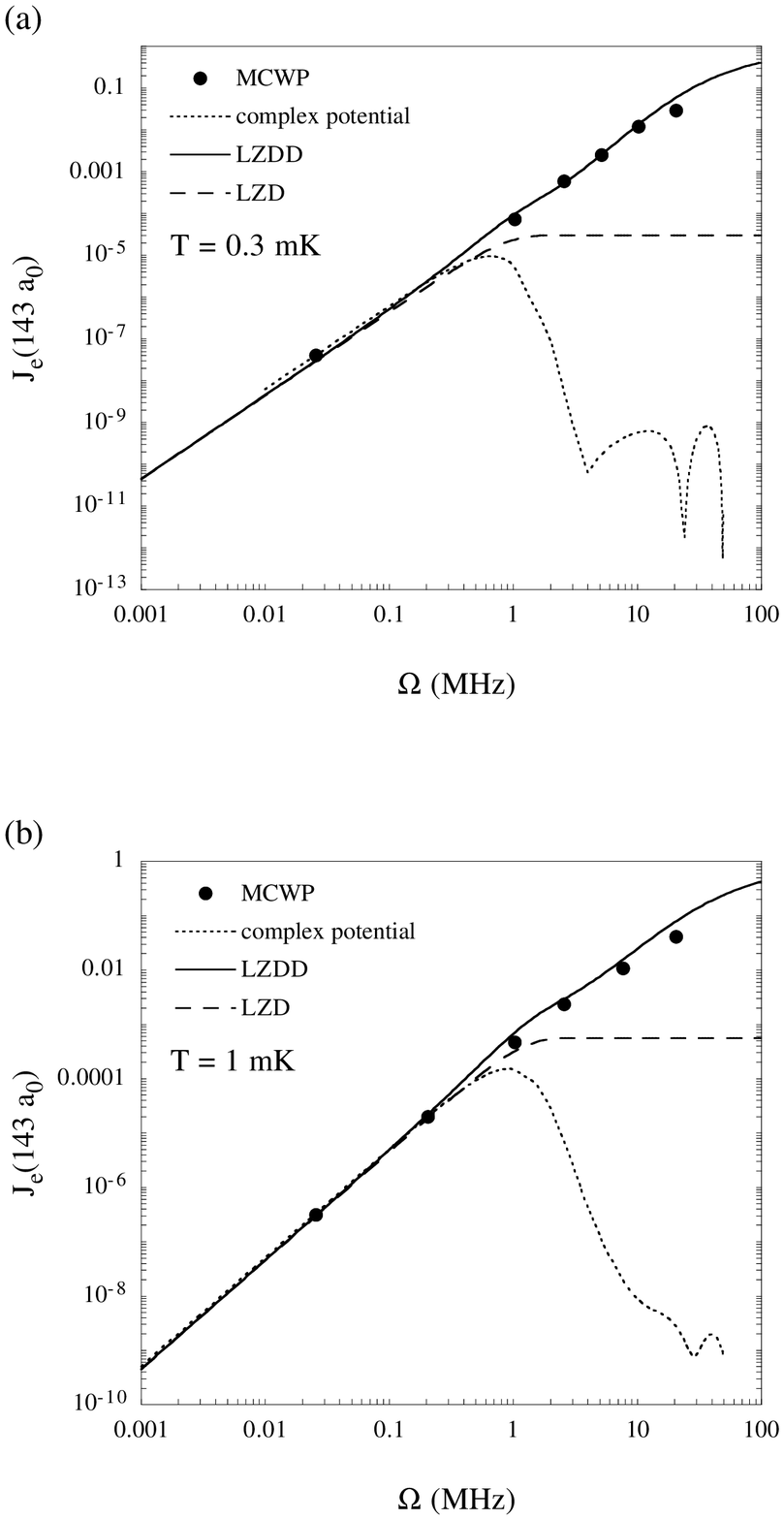}}
\caption[f5]{The excited state flux $J_e$ at $R=143\ a_0$ as a function of the
laser-induced coupling $\Omega$, calculated using the MCWP, complex
potential, LZD and LZDD methods. Here $\Delta=\gamma_{\rm at}$, and the other
parameters are as indicated in the figure.\label{MCWP-Others}}
\end{figure}

We compare the complex potential method and the Landau-Zener approaches to the
MCWP results in Fig.~\ref{MCWP-Others}. The LZD method saturates when the
Landau-Zener excitation probability $P_{\rm LZ}$ becomes unity; until then all
methods seem to agree well. However, beyond the saturation of the excitation
the complex potential approach fails utterly. One should note that the
saturation of the dynamical excitation is not the same as the saturation of
the atomic excitation (also, the atomic excitation saturates to 1/2, but the
dynamical excitation to unity). The complex potential method fails utterly
when one approaches the saturation limit. The LZDD method agrees well with the
MCWP results.

\begin{figure}
\centerline{\psfig{width=200mm,file=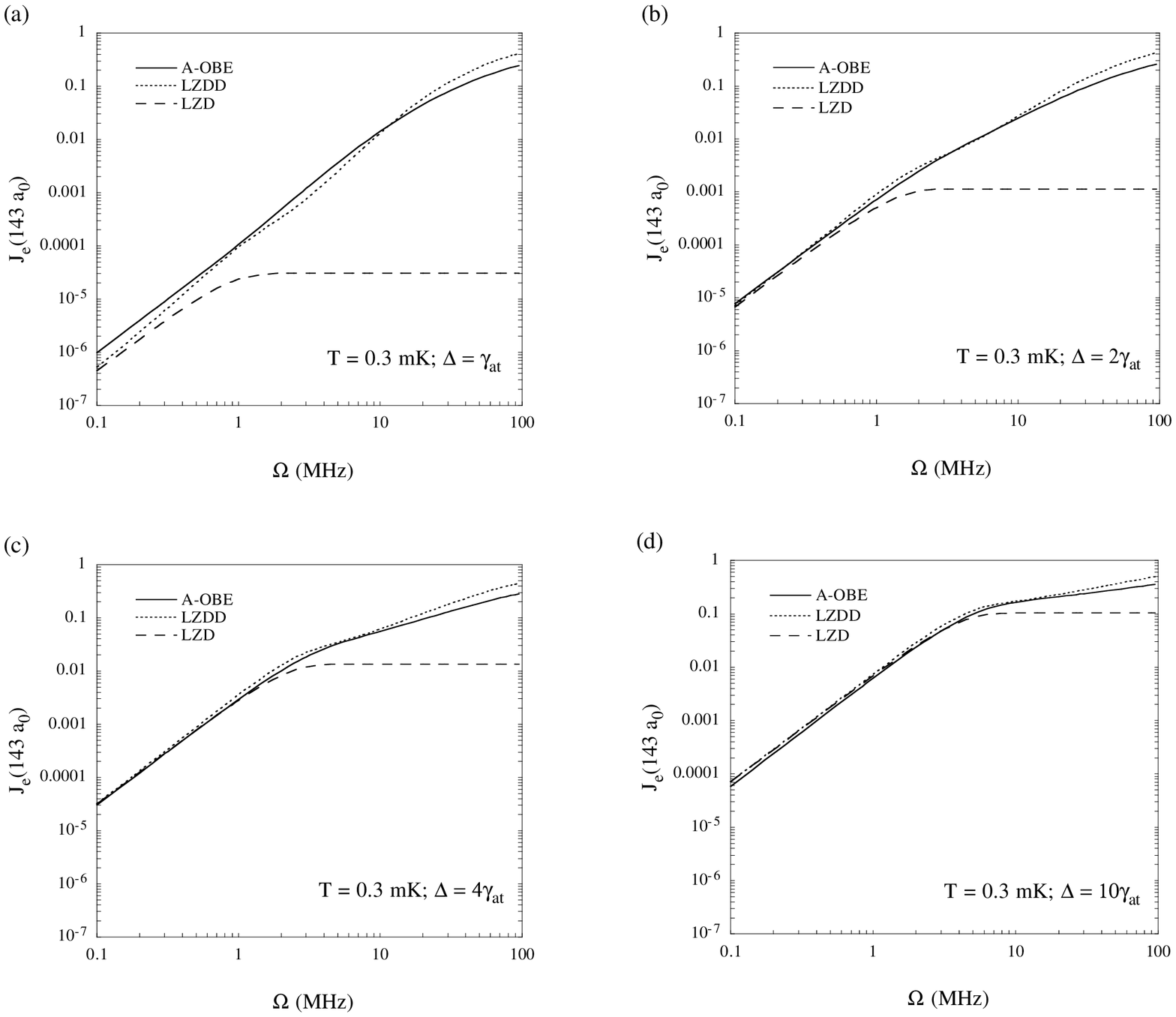}}
\caption[f6]{The excited state flux $J_e$ at $R=143\ a_0$ as a function of the
laser-induced coupling $\Omega$, calculated using the A-OBE, LZD and LZDD
methods for $T=0.3$ mK. \label{t03}}
\end{figure}

\begin{figure}
\centerline{\psfig{width=200mm,file=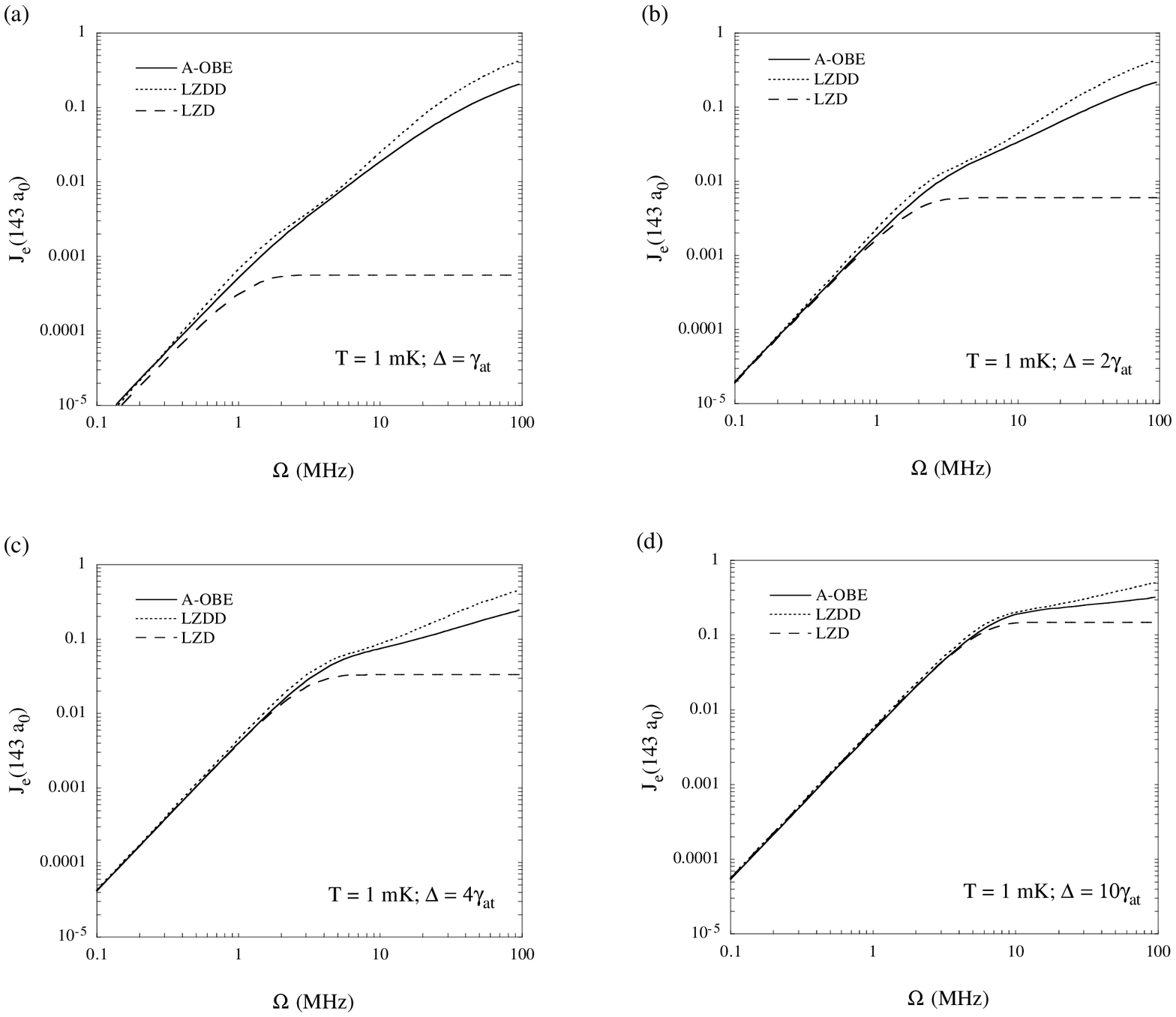}}
\caption[f7]{The excited state flux $J_e$ at $R=143\ a_0$ as a function of the
laser-induced coupling $\Omega$, calculated using the A-OBE, LZD and LZDD
methods for $T=1.0$ mK. \label{t1}}
\end{figure}

We have used the A-OBE and LZDD method to calculate the flux for various
detunings, and the results are given in Figs.~\ref{t03} and \ref{t1}. As can be
expected, the saturation moves to larger $\Omega$ when $\Delta$ increases. In
Fig.~\ref{t03}(a) we start to see the signs of the failure of the Landau-Zener
model at small $T$ (and small $\Delta$). 

\section{Conclusions}\label{final}

In this paper we have derived the adiabatic optical Bloch equations. 
When applied to the standard two-state model for loss of laser-cooled atoms 
from electromagnetic traps, these equations prove to be a fast and adequately 
accurate method to predict probabilities to reach any internuclear distance on
the excited quasimolecule state. The A-OBE results match very well with the 
MCWP results, and also agree with the results from the qualitative LZDD model.
The latter agreement is surprisingly good, but this can be attributed to a
fortuitous definition of the edge of the interaction region.

Although the A-OBE method is a useful tool, we are still far from having a 
proper general treatment of trap loss at small detunings. Firstly, the bound 
state structure of the attractive excited state is not taken into account. At
small detunings the vibrational states associated with the attractive molecular
potential overlap strongly and at strong fields they are also power broadened.
In the language of time-dependent approach, the first ``vibration" of the
quasimolecule dominates over all the others. We can remedy the situation to
some extent by applying the single passage result to multiple passages, as has
been done e.g. in Ref.~\cite{Thad}.

Secondly, the A-OBE method does not allow for momentum change; the motion on
the ground state is given by the ground state velocity factor. When the excited
state population of the quasimolecule decays back to the ground state, its
kinetic energy distribution has been affected by the acceleration on the 
excited state. If this population is excited by e.g. another laser, this 
excitation depends on the new kinetic energy distribution. This effect is
important in the cases of radiative heating~\cite{HSB94} and the radiative
enhancement~\cite{Suominen96b,Gould}. Currently only the MCWP method can 
handle the kinetic energy changes correctly~\cite{HSB94}. 

Thirdly, the velocity factors diverge at classical turning points. Thus the 
A-OBE method is not capable of treating the case of optical 
shielding~\cite{Suominen,Suominen95}, which involves excitation to a repulsive
quasimolecule potential by a blue-detuned laser. This problem might be solved
by replacing the WKB wave function by a proper Airy function solution in our
derivation of the A-OBE method~\cite{Child}.

Finally, we have given here as an example only the case of one ground state 
and one excited state. In reality we have several states involved in the 
cold collision
process, e.g., the various partial waves and hyperfine states. In such a 
situation it is not so easy to write out the transformation to the adiabatic
representation in an analytic form. However, there are various methods for 
doing the change of basis numerically.  A problem may arise from the fact
that we had to redefine the velocity factors using physical arguments---it
is not obvious if such a redefinition in a multistate case would be as
straightforward and clear as in the two-state case. In any case, the A-OBE 
method should be capable of 
handling adequately the case of several partly overlapping strong crossings 
in a system of close-lying states. In such a situation the Landau-Zener 
methods are expected to fail---the A-OBE method can help in testing the 
validity of the Landau-Zener methods in nontrivial crossing situations.

For small detunings the Condon point moves to very large distances and the 
quasimolecule potential does not have the simple $1/R^3$
dependence any more. Furthermore, the retardation effects become important, 
and then the quasimolecular line widths become $R$ dependent even in the 
diabatic representation. The A-OBE method should be very useful in studying
these situations, as it is much faster than the MCWP method when
exploring a wide range of cases with varying laser parameters ($\Delta$ and 
$\Omega$) and quasimolecule potentials ($C_3$, $l$).

Despite some limitations the A-OBE method is a clear leap forward
in treating theoretically the cold collisions in light fields. The discussion
above, however, should be taken as a warning against trusting the method 
too blindly. The results given in this paper and in our previous 
report~\cite{Rapid} indicate nevertheless that the method is very good in 
predicting the behaviour of trap loss due to near-resonant light.

\acknowledgements

This work was supported in part by grants from the U.S.-Israel Binational
Science Foundation and the Office of Naval Research.  K.B. and K.-A. S. thank
the U.K. EPSRC for financial support. K.-A. S. thanks the Academy of Finland
for financial support. The authors also thank Fred Mies for useful comments.


\begin{references}

\bibitem{PAreview} P. D. Lett, P. S. Julienne, and W. D. Phillips,
                   Ann. Rev. Phys. Chem. {\bf 46}, 423 (1995).

\bibitem{Smith}  P. S. Julienne, A. M. Smith, K. Burnett,
                 Adv. At. Mol. Opt. Phys. {\bf 30}, 141 (1993).

\bibitem{Suominen} K.-A. Suominen, 
                   J. Phys. B {\bf 29}, 5981 (1996).

\bibitem{experiments} T. Walker and P. Feng, Adv. At. Mol. Opt. Phys. {\bf
                      34}, 125 (1994);
                      J. Weiner, Adv. At. Mol. Opt. Phys. {\bf 35}, 45 (1995).

\bibitem{densmat} J. von Neumann, {\it Mathematical Foundations of Quantum
                  Mechanics} (Princeton  University Press, Princeton, 1955);
                  L.~van Hove, Physica {\bf 21}, 517 (1955);
                  E. B. Davis, {\it Quantum Theory of Open Systems} (Academic
                  Press, London, 1976); 
                  L. Allen and J. H. Eberly, {\it Optical Resonance and Two
                  Level Atoms}  (Dover, New York, 1987); 
                  R. Alicki and K. Lendi, {\it Quantum Dynamical Semigroups and
                  Applications} (Springer, New York, 1987);
                  H. Carmichael, {\it An Open Systems Approach to Quantum
                  Optics} (Springer, Berlin, 1993);
                  S. Stenholm and M. Wilkens, 
                  Contemp. Phys. {\bf 38}, 257 (1997).

\bibitem{Lai93} W. K. Lai, K.-A. Suominen, B. M. Garraway, and S. Stenholm,
                Phys. Rev. A {\bf 47}, 4779 (1993).

\bibitem{SG93}  K.-A. Suominen and B. M. Garraway,
                Phys. Rev. A {\bf 48}, 3811 (1993).

\bibitem{GP} A. Gallagher and D. E. Pritchard, Phys. Rev. Lett. {\bf 63}, 
             957 (1989).
               
\bibitem{JV} P. S. Julienne and J. Vigu\'e, Phys. Rev. A {\bf 44}, 4464 (1991).

\bibitem{JSB} P. S. Julienne, K.-A. Suominen, and Y. B. Band, Phys. Rev. A {\bf
              49}, 3890 (1994).

\bibitem{HSB94} M. J. Holland, K.-A. Suominen, and K. Burnett,
                Phys. Rev. Lett. {\bf 72}, 2367 (1994);  Phys. Rev. A
               {\bf 50}, 1513 (1994).

\bibitem{Suominen94} K.-A. Suominen, M. J. Holland, K. Burnett, and
                     P. S. Julienne, Phys. Rev. A {\bf 49}, 3897 (1994).

\bibitem{Suominen95} K.-A. Suominen, M. J. Holland, K. Burnett, and
                     P. S. Julienne, Phys. Rev. A {\bf 51}, 1446 (1995).

\bibitem{BJ} Y. B. Band and P. S. Julienne, Phys. Rev. A {\bf 46}, 330 (1992).

\bibitem{TB} I. Tuvi and Y. B. Band, J. Chem. Phys. {\bf 99}, 9697 (1993).

\bibitem{BVT} H. M. J. M. Boesten, B.J. Verhaar, and E. Tiesinga, Phys. Rev. 
              A {\bf 48}, 1428 (1993).
               
\bibitem{Reginaldo} R. Napolitano, J. Weiner, P. S. Julienne, and C. J. 
                    Williams, Phys. Rev. Lett. {\bf 73}, 1352 (1994).

\bibitem{Rapid} Y. B. Band, I. Tuvi, K.-A. Suominen, K. Burnett, and P. S.
                Julienne, Phys. Rev. A {\bf 50}, R2826 (1994).

\bibitem{Arimondo} A. Fioretti, J. H. M{\"u}ller, P. Verkerk, M. Allegrini,
                   E. Arimondo, and P. S. Julienne, Phys. Rev. A
                   {\bf 55}, R3999 (1997).

\bibitem{half} F. H. Mies and P. S. Julienne, J. Chem. Phys. {\bf 80}, 2526 
               (1984); 
               Y. B. Band and F. H. Mies, J. Chem. Phys. {\bf 88}, 2309 
               (1988); 
               R. L. Dubs, P. S. Julienne, and F. H. Mies, 
               J. Chem. Phys. {\bf 93}, 8784 (1990).

\bibitem{BECPRL} K. Burnett, P. S. Julienne, and K.-A. Suominen,
                 Phys. Rev. Lett. {\bf 77}, 1416 (1996).

\bibitem{JNIST} P. S. Julienne, NIST J. Res. {\bf 101}, 487 (1996).

\bibitem{DCM92} J. Dalibard, Y. Castin, and K. M{\o}lmer,
                Phys. Rev. Lett. {\bf 68}, 580 (1992);
                Y. Castin, K. M{\o}lmer, and J. Dalibard,
                J. Opt. Soc. Am. B {\bf 10}, 524 (1993).

\bibitem{Loudon} R.\ Loudon, {\it The Quantum Theory of Light, 2nd ed.}
                 (Oxford University Press, Oxford, 1983).

\bibitem{invimb} S. J. Singer, K. F. Freed, and Y. B. Band, 
                 J. Chem. Phys. {\bf 77}, 1942 (1982).

\bibitem{Thad} M. G. Peters, D. Hoffmann, J. D. Tobiason, and T. Walker,
               Phys. Rev. A {\bf 50}, R906 (1994).

\bibitem{Suominen96b} K.-A. Suominen, K. Burnett, P. S. Julienne, M. Walhout,
                      U. Sterr, C. Orzel, M. Hoogerland, and S. L. Rolston,
                      Phys. Rev. A {\bf 53}, 1678 (1996).

\bibitem{Gould} V. Sanchez-Villicana, S. D. Gensemer, and P. L. Gould,
                Phys. Rev. A {\bf 54}, R3730 (1996).

\bibitem{Child}  M. S. Child, {\it Semiclassical Mechanics with Molecular
                 Applications} (Oxford University Press, Oxford, 1991).

\end{references}
\end{document}